\documentclass[aps,prl,twocolumn,secnumarabic,amssymb, amsmath,nobibnotes,showpacs]{revtex4-1}

\usepackage{amssymb,amsmath,amscd}
\usepackage{graphicx,calc,epsfig,pstricks,bbm}
\usepackage{tikz}
\usepackage{enumerate}
\expandafter\ifx\csname package@font\endcsname\relax\else
 \expandafter\expandafter
 \expandafter\usepackage
 \expandafter\expandafter
 \expandafter{\csname package@font\endcsname}%
\fi

\newcommand {\be}{\begin{equation}}
\newcommand {\ee}{\end{equation}}
\newcommand{\ba}{\begin{array}{c}}
\newcommand{\ea}{\end{array}}

\begin{document}
\title{Improved regularization from Quantum Reduced Loop Gravity}%

\author{Emanuele Alesci$^{1}$, Francesco Cianfrani$^{2}$}%
\affiliation{$^{1}$
SISSA, Via Bonomea 265, 34126 Trieste, Italy and INFN Sez. Trieste.\\
$^{2}$Institute
for Theoretical Physics, University of Wroc\l{}aw, Pl.\ Maksa Borna
9, Pl--50-204 Wroc\l{}aw, Poland.}
\date{\today}%

\begin{abstract} 
The choice of the regularization scheme in Loop Quantum Cosmology (LQC) is crucial for the predicted phenomenology. We outline how the improved scheme can 
be naturally realized in Quantum Reduced Loop Gravity, describing the Universe as an ensemble of microstates labeled by different graphs. The new effective dynamics presents corrections to LQC, which do not significantly affect the bouncing scenario for the most relevant kinds of matter fields in cosmology ($w\leq 1$). 
\end{abstract}


\maketitle

The main arena for a quantum theory of gravity is cosmology. The general expectation is that Quantum Gravity tames the initial cosmological singularity and the hope is that the modifications to the Friedman dynamics could induce some potentially testable effects on the behavior of perturbations (to be constrained with the measurement of cosmic microwave background radiation spectrum and, in perspective, of the spectrum of cosmological gravitational waves).   

Among other approaches, Loop Quantum Gravity (LQG) \cite{Rovelli:2004tv,Thiemann:2007zz,Ashtekar:2004eh} is the one closest to the spirit of General Relativity, {\it i.e.} to the geometric interpretation of the gravitational field, and it tries to recombine in the most straightforward way the quantization procedures and a dynamical geometry. 
Important achievements have been obtained in the standard cosmological implementation of LQG, namely Loop Quantum Cosmology (LQC). LQC \cite{Bojowald:2011zzb,Ashtekar:2011ni} is a minisuperspace quantization in which the tools of loop quantization are adopted. This means that the phase space of gravity is reduced according with the symmetries of FRW models and the resulting reduced phase space is then quantized. 
The model succeeds in providing a singularity-free description of the Universe, since the initial singularity is replaced with a bounce occurring when matter energy density reaches Planckian values \cite{Singh:2006im,Ashtekar:2006rx}. 

There is one important technical assumption behind the derivation of the bounce in LQC. The quantization procedure results in a polymer quantum representation applied to the reduced phase space. Polymer representation \cite{Halvorson,Ashtekar:2002sn,Corichi:2007tf} is inequivalent to Schr\"odinger representation and it describes models in which configuration variables or momenta are discrete. The scale of such a discretization is entailed into one parameter, the so-called polymer parameter. The choice of the value of such parameter, labeling nonequivalent quantum theories, is not inherent to LQC, rather it is made importing the minimum area eigenvalue from LQG and requiring that quantum corrections are relevant only when matter energy density reaches Planckian values. This results in a preferred choice, known as the improved regularization scheme, corresponding to a polymer parameter proportional to the inverse scale factor. 

Recently, it has been proposed an alternative derivation of the cosmological sector of LQG, Quantum-Reduced Loop Gravity (QRLG) \cite{Alesci:2012md,Alesci:2013xd}, based on reducing the kinematical Hilbert space of LQG according with some gauge-fixing conditions and implementing the reduction to homogeneity just on the Hamiltonian. QRLG stands at an intermediate level between full LQG and LQC, since it preserves some of the basic LQG structures (a discrete graph and intertwiners), but in a simplified context in which computations can be performed analytically, reproducing in the semiclassical limit the kind of effective Hamiltonian of LQC \cite{Alesci:2014uha,Alesci:2014rra,Alesci:2015nja}. 
This framework reveals how the origin of the polymer parameter is deeply rooted in LQG, since it equals the third root of the inverse number of nodes of the graph at which quantum states are based. However, this number is fixed in the current version of QRLG (based on a non-graph changing Hamiltonian), thus it cannot be identified with the polymer parameter in improved regularization. A natural extension of this framework, that would take a varying number of nodes into account, would be the implementation of a graph changing Hamiltonian that could relate the growth of the scale factors to the creation of new nodes.

In this work, we will outline instead how it is possible to derive such a regularization scheme even in QRLG with a non-graph changing Hamiltonian. The idea is to construct a density matrix with microstates having different number of nodes and quantum numbers such that they realize the same macroscopic semiclassical configuration. Given an algebraic graph with $N_{max}$ cells, there are $\binom{N_{max}}{N}$ different realizations of microstates with $N$ nodes. Hence, the resulting matrix density is a sum over states with $N$ nodes weighted by the binomial coefficient $\binom{N_{max}}{N}$. We compute the expectation value of the effective Hamiltonian via a saddle-point expansion, approximating the weight with a Gaussian. At the leading order, one gets LQC effective Hamiltonian with the improved regularization. We also test whether the next-to-the-leading order term of the saddle-point expansion can affect the result. Treating this term as a perturbation, we demonstrate how the solutions of the equations of motion do not depart significantly from those of LQC, at least for the most relevant cases in cosmology, {\it i.e.} in the presence of matter fields with $w\leq 1$.

\paragraph{LQC-} The flat FRW model is described by the line element
\be
ds^2=dt^2-a^2(t)\left(dx^2+dy^2+dz^2\right)\,,
\ee
where $a$ denotes the scale factor and it is a function of time only. In LQC, the phase space is 
described by the following coordinates 
\be
p=\ell_0^{-2}\,a^2\qquad c=\gamma\ell_0^{-1}\,\dot{a}\,,\label{pc}
\ee
$\gamma$ being Immirzi parameter, while $\ell_0$ denotes the considered fiducial length scale. The only nonvanishing Poisson brackets are 
\be
\{p,c\}=\frac{8\pi G\gamma}{3}\,,
\ee
and the Hamiltonian is
\begin{equation}
H=\frac{1}{8\pi G}\mathcal{S}\qquad \mathcal{S}=\frac{3}{\gamma^2}\sqrt{p}c^2-8\pi G\, p^{3/2}\rho=0\label{Sh}\,,
\end{equation}
where $\mathcal{S}$ denotes the scalar constraint 
in the presence of a matter field described by an energy density $\rho$, which generically is a function of matter degrees of freedom and of $p$. 

In LQC, the effective dynamics is obtained by making the following replacement in \eqref{Sh} 
\be
c\rightarrow \sin(\bar\mu c)/\bar\mu\,,\label{repl}
\ee
which is a standard procedure in polymer quantization. The polymer parameter $\bar\mu$ is fixed as follows 
\be
p\bar\mu^2=\Delta=4\pi\sqrt{3}\gamma \ell^2_{pl}\,\label{barmu}
\ee
and this is a choice of a particular (so-called improved) regularization scheme, coming from the requirement that the minimal area gap in LQC coincides with those of full LQG. The resulting effective Hamiltonian reads
\begin{equation}
H=\frac{3}{8\pi G\gamma^2}\sqrt{p}\left(\frac{\sin(\bar\mu c)}{\bar\mu}\right)^2-p^{3/2}\,\rho(p)=0\,,\label{qrlghiso} 
\end{equation}
and it is constrained to vanish. From the equation of motion for $p$, by using \eqref{pc} and \eqref{qrlghiso}, one can infer the modified Friedmann equation
\begin{equation}
\left(\frac{\dot{a}}{a}\right)^2=\frac{8\pi G}{3}\,\rho\left(1-\frac{\rho}{\rho_{cr}}\right)\qquad \rho_{cr}=\frac{3}{8\pi G\gamma^2\Delta}\,,
\end{equation}
where the critical energy density $\rho_{cr}$ is of Planckian energy density order.
 
When the energy density $\rho$ reaches the critical energy density the Universe experiences a bounce and the initial singularity is avoided.

Generically, one can consider other choices for the parameter $\bar\mu$, but they usually provides some un-wanted features, as for instance a critical energy density depending on $p$, which can induce quantum gravity effects also at much smaller densities than the Planckian one. In what follows, we will outline how to derive the improved regularization scheme in the context of QRLG.

\paragraph{Summing over graphs in QRLG-} QRLG is an alternative derivation of the cosmological sector of LQG with respect to LQC. It is based on implementing some gauge-fixing conditions directly in the kinematical Hilbert space of the full theory. The resulting model is described in terms of cubical graphs, similarly to full theory, but despite that it can be treated analytically. In particular, each quantum state in QRLG is described by some $U(1)$ quantum numbers at links and some nontrivial intertwiners at nodes. We can think at the graph as a collection of cubical cells and the collective classical variables $p$ and $c$ are now obtained from those at a single cell via a sum over all cells.

Semiclassical states \cite{Alesci:2014rra} have been constructed in the homogeneous and isotropic case by peaking the variables of all cells around the same expectation values. Each of these states is labeled by the total number $N$ of the nodes of the graph at which it is based and the single-cell variables $j$ and $\theta$ around which it is peaked, $j$ being $U(1)$ quantum number and $\theta$ denoting the connection along a link. We denote such state by $|N,j,\theta\rangle$. Both the graph and single-cell variables determine the values of classical phase space coordinates as follows
\be
p=8\pi\gamma \ell_P N^{2/3} j\qquad c=N^{1/3}\theta\,,\label{pqrlg}
\ee  
By retaining only the part of the Hamiltonian generating the dynamics of the homogeneous modes of the metric $H^{grav}$, it has been derived in the semiclassical limit the following effective Hamiltonian for isotropic and homogeneous configurations \cite{Alesci:2015nja}
\be
\langle N,j,\theta|\hat{H}^{grav}|N,j,\theta\rangle=\frac{3}{8\pi G\gamma^2}\sqrt{p}N^{2/3}\sin^2(N^{-1/3} c)\,.\label{hgrav}
\ee
The expression above resembles the gravitational part of \eqref{qrlghiso}, except for the choice of the polymer parameter, which we now call $\mu_0$ and it is related to the total number of nodes $N$ as follows (the same identification occurs in lattice-refined LQC \cite{Bojowald:2006qu} and Group Field Theory cosmology \cite{Gielen:2014uga,Oriti:2016qtz})
\be
\mu_0=N^{-1/3}\,.
\ee
The point is that we cannot identify $\bar\mu$ and $\mu_0$, since the expression for the Hamiltonian constraint and the adopted semiclassical tools are based on the assumption that the Hamiltonian itself is not graph-changing, {\it i.e.} that the number of nodes $N$ is constant. A constant polymer parameter $\mu_0$ has been considered at early stages of LQC developments \cite{Ashtekar:2006uz}, but it was then abandoned for the $\bar\mu$ scheme. However, in connection with QRLG, the cosmological implications of $\mu_0$ regularization scheme has been recently reconsidered in \cite{Cianfrani:2015oha}. 

We are now going to discuss how to pass from the $\mu_0$ to the improved regularization via a summation over graphs. 

Our analysis starts from recognizing how a given classical configuration, {\it i.e.} a point in the reduced phase space $(p,c)$, is associated to many different semiclassical states as soon as many graphs are considered. This is clear from \eqref{pqrlg}, where the two phase space coordinates are determined by three microscopic variables, namely the quantum number $j$ and $\theta$ at single links and the total number of nodes $N$. Hence, a configuration with a given $p$ and $c$ can be realized in many ways, changing $N$, $j$ and $\theta$ such that $N^{2/3} j$ and $N^{1/3}\theta$ are constant. 

Hence, let us construct a density matrix as a sum of states having the same $p$ and $c$
\be
\rho_{p,c}=\sum_{N}c_{N} \, |N,j(p,N),\theta(c,N)\rangle\langle N,j(p,N),\theta(c,N)|\,,\label{mix}
\ee
where $j(p,N)$ and $\theta(c,N)$ are obtained by inverting the relations \eqref{pqrlg} and the sum extends over $N$ only. The coefficient $c_N$ modulo a normalization equals the multiplicity of each $N$, {\it i.e.} the number of different graphs with $N$ nodes one can construct on an algebraic graph having $N_{max}$ cells (similarly to the idea behind Algebraic Quantum Gravity \cite{Giesel:2006uj}). Given $p$, $N$ ranges from $1$ (there must be at least one node) to a maximum value $N_{max}$ corresponding to a minimum nonvanishing value for $j$ which we call $j_0$ (we expect $j_0=1/2$ but we prefer to keep it arbitrary). This gives
\be
c_N=\frac{1}{2^{N_{max}}} \binom{N_{max}}{N}\,,
\ee
where $N_{max}$ is determined by the following relation
\be
p=8\pi\gamma\ell_P^2 N^{2/3}_{max} j_0.   \label{pnmax}
\ee
If we evaluate the expectation value of the Hamiltonian over $\rho$ we get
\be
Tr(\rho_{p,c} H)=\frac{3}{8\pi G\gamma^2}\sqrt{p}\,f(c,N_{max})\,,\label{geh}
\ee
where
\be
f(c,N_{max})=\frac{1}{2^{N_{max}}}\sum_{N=1}^{N_{max}} \binom{N_{max}}{N}\,N^{2/3} \sin^2(c/N^{1/3})\,.
\ee
For physical reasons we can formally take the limit $N_{max}\rightarrow \infty$, in which case we can approximate the binomial coefficient with a Gaussian function centered around $N_{max}/2$ with a variance $\sigma^2=N_{max}/4$ and the sum with an integral, so getting
\be
f(c,N_{max})\sim \int \frac{1}{\sqrt{\pi N_{max}/2}}\,e^{-\frac{(N-N_{max}/2)^2}{N_{max}/2}}\,N^{2/3} \sin(c/N^{1/3})\,dN\,,\label{fsp}
\ee
which via a saddle point expansion at the leading order provides just the integrand evaluated at the center of the Gaussian in $N=N_{max}/2$, {\it i.e.}
\be
f(c,N_{max})\sim \frac{\sin^2(\tilde\mu c)}{\tilde\mu^2}\,, \label{fcom0}
\ee 
where we introduced $\tilde\mu=(N_{max}/2)^{-1/3}$ for which 
\be
p\tilde\mu^2=\tilde\Delta\qquad \tilde\Delta=(2)^{2/3}8\pi\gamma\ell^2_{pl}j_0\,.\label{mu}
\ee
It is worth noting how by substituting \eqref{fcom0} in \eqref{geh}, one ends up with the same kind of polymer-like replacement \eqref{repl} as with improved regularization. In fact, $\bar\mu$ and $\tilde\mu$ have the same $p$-dependence and they just differ by a constant factor ($\Delta$ in \eqref{barmu} vs $\tilde\Delta$ in \eqref{mu}). Therefore, at the leading order of the saddle-point expansion in \eqref{fsp}, the effective gravitation Hamiltonian is essentially that of improved regularization. Next-to-the-leading orders are suppressed by a factor of order $N_{max}^{-2}$ and we expect them to be negligible. 
In what follows, we will test this hypothesis, but we first have to solve the equations of motion at the leading order in the presence of a classical matter field described by an energy density $\rho=\rho_0 p^{-\alpha}$. From \eqref{geh}, the effective Hamiltonian reads
\be
H_{eff}=-\frac{3}{8\pi G\gamma^2}\sqrt{p}\,f(c,N_{max})+\rho_0 \, p^{3/2-\alpha}\,.\label{modh}
\ee
The equations of motion can be rewritten as follows by introducing the variable $x=\tilde\mu c$ and the constant $K=\frac{8\pi G\gamma^2\tilde\Delta \rho_0}{3}$
\begin{align}
&\dot{p}=-\frac{1}{\gamma\tilde\Delta^{1/2}}\, p \,\sin(2x)\label{p.0}\\ 
&\dot{x}=\frac{K\alpha}{\gamma\tilde\Delta^{1/2}} \, p^{-\alpha} \label{x.0}
\end{align}
where we used the condition $H_{eff}=0$, which is equivalent to
\be
\sin^2x=\frac{K}{\rho_0}\,\rho\,.\label{h0}
\ee
Let us denote by $\bar{p}$ and $\bar{x}$ the analytic solutions of the system of equations \eqref{p.0}, \eqref{x.0} and \eqref{h0} explicitly given by 
\begin{align}
&\cos(2\bar{x}) = 1-2K \bar{p}^{-\alpha}\,,\label{solx}\\
&\bar{p}^{\alpha}(t)= K+ \left[\sqrt{\bar{p}^{\alpha}_i-K}-\frac{\sqrt{4K}\alpha}{\gamma\Delta^{1/2}} (t-t_i)\right]^{2}\,.\label{solp}
\end{align}
It is worth noting how $\bar{p}(t)$ has a minimum, corresponding to its value at the bounce $\bar{p}_B=K^{1/\alpha}$, which is reached for $t=t_B$ such that the second term on the right-hand side of \eqref{solp} vanishes.

We can now discuss the impact of next orders in the saddle point expansion. 
The inclusion of the next-to-the leading order term in \eqref{fsp} provides the following modification to \eqref{fcom0} 
\be
f(c,N_{max})\sim \frac{\sin^2(\tilde\mu c)}{\tilde\mu^2}+\frac{1}{9}\tilde\mu^3 c^2 \cos(2\tilde\mu c)\,, \label{fcom}
\ee
and the new equations of motion become
\begin{align}
\dot{p}=&-\frac{1}{\gamma\tilde\Delta^{1/2}}\, p \,\sin(2x)\bigg[1-\frac{2}{9}\lambda\,\left(\frac{\bar{p}_B}{p}\right)^{3/2}x^2 \nonumber\\
&\hspace{2.5cm}+ \frac{2}{9}\lambda\,\left(\frac{\bar{p}_B}{p}\right)^{3/2}x\,\cot(2x)\bigg]\label{p.}\\ 
\dot{x}=&-\frac{8\pi G\gamma\tilde\Delta^{1/2}}{3} \, p\,\frac{d\rho}{dp}-\lambda\frac{1}{6\gamma\tilde\Delta^{1/2}}\,\left(\frac{\bar{p}_B}{p}\right)^{3/2}x^2\,\cos(2x) \label{x.}\\
\sin^2x&+\frac{\lambda}{9}\, \left(\frac{\bar{p}_B}{p}\right)^{3/2}x^2\cos(2x)=\frac{8\pi G\gamma^2\tilde\Delta}{3}\,\rho\,,\label{h}
\end{align}
where we introduced the parameter $\lambda$
\be
\lambda=\left(\frac{\tilde{\Delta}}{\bar{p}_B}\right)^{3/2}=\frac{1}{N_{max}^{1/2}(\bar{p}_B)}\ll 1\,.\label{lambda}
\ee
We want to study to which extend we can approximate the solutions to the system of equations above with \eqref{solx} and \eqref{solp}. In this respect, we perturb \eqref{p.}, \eqref{x.} and \eqref{h} as follows: we write 
\be
p=\bar{p}+\lambda\,\delta p\qquad x=\bar{x}+\lambda\,\delta x\,,
\ee
where $\lambda\,|\delta p|\ll |\bar{p}|$ and $\lambda\,|\delta x|\ll |\bar{x}|$, and we study the resulting dynamic equations for $\delta p$ and $\delta x$, treating the terms proportional to $\lambda$ in \eqref{p.}, \eqref{x.} and \eqref{h} as perturbations. In other words, we substitute $p=\bar{p}$ and $x=\bar{x}$ in $\lambda$-terms and we treat them as sources at the first order of the perturbative expansion. This is legitimate, since we know these contributions to be negligible today and we want just to investigate whether they can influence the dynamics of perturbations up to invalidate the perturbative expansion and affect the bouncing scenario. In particular, if $\lambda\,|\delta x|\ll |x|$ and $\lambda\,|\delta p|\ll |p|$ everywhere, our approximation scheme is consistent and additional terms are negligible. On the contrary, if at some point $\lambda\,|\delta x|\sim |x|$ or $\lambda\,|\delta p|\sim |p|$, the perturbative expansion is not well defined and the solutions to \eqref{p.}, \eqref{x.} and \eqref{h} significantly differ from those of LQC, \eqref{solx} and \eqref{solp}. 

This way, we get the following equations for $\delta p$ and $\delta x$ 
\begin{align}
\dot{(\delta p)}=&-\frac{1}{\gamma\tilde\Delta ^{1/2}} \,\sin(2\bar{x})\delta p-\frac{2}{\gamma\tilde\Delta ^{1/2}}\, \bar{p} \,\cos(2\bar{x})\,\delta x+\nonumber\\
&\;\;\frac{2\bar{p}_B^{3/2}}{9\gamma\tilde\Delta^{1/2}\bar{p}^{1/2}}\, \bar{x}^2 \,\sin(2\bar{x})-\frac{2\bar{p}_B^{3/2}}{9\gamma\tilde\Delta^{1/2}\bar{p}^{1/2}}\, \bar{x}\, \cos(2\bar{x})\label{deltap.}\\ 
\dot{(\delta x)}=&-\frac{K\alpha^2}{\gamma\tilde\Delta ^{1/2}}\,\bar{p}^{-\alpha-1} \,\delta p-\frac{\bar{p}_B^{3/2}}{6\gamma\tilde\Delta^{1/2}\bar{p}^{3/2}}\,\bar{x}^2\,\cos(2\bar{x}) \label{deltax.}\\
\frac{1}{\tilde\Delta }\,\bar{p}^{3/2}&\,\sin(2\bar{x})\,\delta x+\frac{\bar{p}_B^{3/2}}{9\tilde\Delta}\,\bar{x}^2\, \cos(2\bar{x})=-\frac{K}{\tilde\Delta }\,\alpha\,\bar{p}^{-\alpha+1/2}\,\delta p\,.\label{deltah0}
\end{align}
We can solve this system of equations, so getting for $\delta x$
\be
\delta x(\bar{p})=\delta x^{LQC}(\bar{p}) - \frac{2^{2-3/(2\alpha)}}{9\alpha}\,\left(\alpha-\frac{3}{2}\right)\,I(\bar{x})\,\left(\frac{\bar{p}_B}{\bar{p}}\right)^{\alpha},
\ee
where 
\begin{align}
I(\bar{x})=\int^{\bar{x}}_{\bar{x}_i} dx'\, \frac{(x')^2\,\cos(2x')}{[1-\cos(2x')]^{2-3/(2\alpha)}}\,,\label{I}
\end{align}
and $\delta x^{LQC}$ describes the behavior of homogeneous perturbations in improved LQC, {\it i.e.} it can be obtained directly from \eqref{p.0}, \eqref{x.0} and \eqref{h0} by varying $x$ and $p$. Hence, we are interested in the term $\delta x-\delta x^{LQC}$ in order to investigate how the dynamics generated by the modified effective Hamiltonian \eqref{modh} differs from that of improved LQC. It is worth noting how for $\alpha=3/2$, which corresponds to matter-dominated Universe, no modification occurs. 

Let us study $\delta x-\delta x^{LQC}$: for $\alpha\geq 0$, which corresponds to $w\geq -1$, the factor $(\bar{p}_B/\bar{p})^{\alpha}$ is at most $1$ at the bounce, thus it cannot provide any sensible enhancement to $\delta x-\delta x^{LQC}$.

As a final step, let us show that the same conclusion holds also for $I(\bar{x})$ at least for $\alpha\leq 3$, {\it i.e.} for the most relevant cases in cosmology. Since the range of variation of $x'$ is bounded by $\bar{x}(\bar{p}_B)=-\frac{\pi}{2}$ and $\bar{x}_i\sim 0$, the only dangerous contribution may come near $\bar{x}_i\sim 0$, since at $\bar{x}'=0$ the integrand in \eqref{I} diverges. We can approximate the integral taking the leading order term of the Taylor expansion of the integrand close to $0$, {\it i.e.}
\be
I(\bar{x})\sim 2^{-2+3/(2\alpha)}\int^{\bar{x}}_{\bar{x}_i} dx'\, \frac{1}{(x')^{2-3/\alpha}}\,.
\ee
which gives a finite contribution for $\alpha<3$. For $\alpha=3$, which corresponds to a massless non-interacting scalar field, a similar calculation gives using \eqref{solx}
\be
I(\bar{x})\sim -2^{-3/2}\,\log(\bar{x}_i)\sim -3\times 2^{-5/2}\,\log\left(\frac{\bar{p}_B}{\bar{p}_i}\right)\,,\label{IB}
\ee
which provides an enhancement to $\delta x-\delta x^{LQC}$, but only through a logarithmic factor. An estimate of $\bar{p}_B/\bar{p}_i$ can be given from \eqref{solp} 
\be
\left(\frac{\bar{p}_i}{\bar{p}_B}\right)^{3/2}\sim \frac{3\sqrt{4}}{\gamma\tilde\Delta^{1/2}} (t_i-t_B)\sim 10^{60} \,,
\ee
where we approximate $t_i-t_B$ with the age of the Universe, $10^{28} cm$, and assume $\gamma\tilde\Delta^{1/2}\sim \ell_{pl}$. Hence, the logarithmic factor in \eqref{IB} provides at most a factor of order $100$, which does not substantially affect the perturbative expansion for $\lambda\ll 10^{-2}$ \footnote{One can easily show using \eqref{deltah0} that similar conclusions hold for $\delta p$.}.

Therefore, during the entire history of the Universe, the effective semiclassical dynamics of the mixed state \eqref{mix} is that of improved LQC for $\alpha< 3$, which corresponds to $w< 1$, as soon as the condition \eqref{lambda} holds. For $\alpha=3$, {\it i.e.} for a massless noninteracting scalar field, the stronger condition $\lambda\ll 10^{-2}$ must be imposed.  

\paragraph{Conclusions-} We demonstrated how the improved regularization ($\bar\mu$) scheme can be derived in QRLG, by constructing the Universe as an ensemble of microstates labeled by different graphs. The improved regularization scheme is expected to emerge from a graph-changing Hamiltonian able to add new nodes as the Universe grows. Here, instead, we derived such a scheme acting with a non-graph changing Hamiltonian. The key point is that even if the Hamiltonian changes only quantum numbers, the weight (taking into account the combinatorial factor counting the number of microstates associated with the same semiclassical configuration) peaks the summation around a scale factor-dependent number of nodes. This effectively gives that the relevant number of nodes contributing to the summation changes in time. 

These achievements outline how QRLG is able to bring the ideas and tools of LQG in a cosmological context, leading to new modified Friedman equation. At the leading order of a saddle point expansion, the effective semiclassical dynamics of LQC in improved regularization emerges, so unveiling for the first time its connection with LQG. We also investigated next-to-leading order terms, showing how they do not significantly affect the bounce, even if they can potentially provide new and completely unexplored modifications to the standard scenario.

{\it Acknowledgment}   
FC is supported by funds provided by the National Science Center under the agreement
DEC-2011/02/A/ST2/00294. 
EA wishes to acknowledge the John Templeton Foundation for the supporting grant \#51876.

\end{document}